\documentclass[epj,nopacs]{svjour}
\pdfoutput=1
\usepackage{amsmath,amssymb,amsfonts,graphicx,cite,multirow,color,algorithmic,algorithm}
\usepackage[utf8]{inputenc}
\usepackage{amsmath}
\usepackage{caption}
\usepackage{subcaption}
\usepackage{hyperref}
\graphicspath{{graphics/}}
\makeatletter
\ifx\input@path\@undefined
\def\input@path{{graphics/}}
\else
\g@addto@macro\input@path{{graphics/}}
\fi
\makeatother

\newcommand{\CV}{C_V}
\newcommand{\ESS}{\texttt{ESS}}

\input{def.sty}

\preprint{LU-TP 19-54\\UWTHPH-2019-12\\MCnet-19-25}

\title{Resampling Algorithms for High Energy Physics Simulations}

\author{Jimmy Olsson\inst{1}, Simon Pl\"atzer\inst{2} and Malin Sj\"odahl\inst{3}}

\institute{Department of Mathematics,
  KTH Royal Institute of Technology,
  Lindstedtsv\"{a}gen 25, 100 44 Stockholm, Sweden 
  \and Particle Physics, Faculty of Physics, University of Vienna,
  Boltzmanngasse 5, 1090 Wien, Austria
  \and Department of Astronomy and Theoretical Physics, Lund
  University, S{\"o}lvegatan 14A, 223\,62 Lund, Sweden}

\date{\today}

\abstract{We demonstrate that the method of interleaved resampling in
  the context of parton showers can tremendously improve the
  statistical convergence of weighted parton shower evolution
  algorithms. We illustrate this by several examples showing
  significant statistical improvement.}

\begin{document}

\maketitle


\section{Introduction}
\label{sec:Introduction}

Event generators are indispensable simulation tools for understanding high
energy collisions in particle physics.  A core part of the event generators
are parton showers where hard scattering events, calculated
perturbatively, are dressed up with more and more partons in an iterative
manner, typically starting from some high energy scale $Q$ and evolving down
towards smaller and smaller scales until an infrared cutoff $Q_0$ is reached.

Branchings within the parton shower evolution occur with a rate
$P(q,z,x)$, which determines the probability density of the dynamic
variables associated to the branching. These are the scale $q$,
additional splitting variables $z$, and additional variables $x$ on
which a certain type of branching depends. The latter can be
collections of discrete and continuous values. Branchings are ordered
in decreasing scale $q$ in the evolution interval $Q$ to $Q_0$.  The
probability for a certain splitting not to occur between the scales
$Q$ and $q$ is given by the so-called Sudakov form factor
$\Delta_P(q|Q,x)$,
\begin{equation}
  \Delta_P(q|Q,x) = \exp\left(-\int_q^Q {\rm d}k\int {\rm d}z\ 
  P(k,z,x) \right)\;.
\end{equation}
The density for the dynamic variables describing an individual
branching which is to occur below a scale $Q$ of a previous
branching (or an initial condition set by the hard process), is then
given by
\begin{multline}
  \frac{{\rm d}S_P (q|Q,z,x)}{{\rm d}q\  {\rm d}z} =
  \Delta_P(Q_0|Q,x)\delta(q-Q_0)\delta(z-z_0) \\ +
  \Delta_P(q|Q,x)P(q,z,x)\theta(Q-q)\theta(q-Q_0), 
\end{multline}
where $\Delta_P(Q_0|Q,x)\delta(q-Q_0)\delta(z-z_0)$ represents events
which did not radiate until they reached the cut-off scale $Q_0$, and
$z_0$ is an irrelevant parameter associating fixed branching variables
to the cutoff scale, where no actual branching will occur. Different
branching types $x_1,\ldots, x_n$, with a total rate given by
$\sum_{i=1}^n P(q,z,x_i)$, can be drawn algorithmically using the
competition algorithm (see {\it e.g.}
\cite{Seymour:1994df,Platzer:2011dq}), which we will not discuss in
more detail in this letter.

An individual splitting kernel $P(q,z,x)$ cannot generally be integrated to an
invertible function, which would allow to draw from ${\rm d}S_P$ using
sampling by inversion, so what is typically done is to find an
overestimate $R(q,z,x)$ (s.t. $R(q,z,x) \geq P(q,z,x)$ for all $q,z,x$),
which can be integrated to an invertible function. Equating the
normalized integrated splitting kernel to a random number $r \in
(0,1)$ and solving for $q$ then generates a new scale $q$ with the
distribution for $R(q,z,x)$, and it can be proved that accepting the
proposed branching variables with probability $P(q,z,x)/R(q,z,x)$
generates the desired density $P(q,z,x)$ instead
\cite{Seymour:1994df,Platzer:2011dq,Buckley:2011ms,Platzer:2011dr,Lonnblad:2012hz}.

In practice, in the case of a positive definite splitting kernel
$P(q,z,x)$ with a known overestimate $R(q,z,x)$, the algorithm to draw
from ${\rm d}S_P$ then is given by Alg.~\ref{algos:sud}.
\begin{algorithm}
  \begin{algorithmic}
    \STATE $Q' \gets Q$
    \LOOP
    \STATE $r\gets ${\bf rnd}
    \IF{$r\le \Delta_R(Q_0|Q',x)$}
    \RETURN $Q_0$
    \ELSE
    \STATE solve $r=\Delta_R(q|Q',x)$ for $q$
    \STATE select $z$ in proportion to $R(q,z,x)$
    \RETURN $q$ with probability $P(q,z,x)/R(q,z,x)$
    \ENDIF
    \STATE $Q'\gets q$
    \ENDLOOP
  \end{algorithmic}
  \caption{\label{algos:sud}The Sudakov veto algorithm, defined using
    an overestimate $R$ such that $R(q, z, x) \geq P(q, z, x)$ for all $q, z, x$.}
\end{algorithm}
This algorithm is statistically perfect in the sense that it produces
the correct distribution using only (positive) unit weights.  In case
of splitting kernels of indefinite sign, or in an attempt to include
variations of the splitting rate $P$ attached to a fixed sequence of
branching variables, weights different from unity are unavoidable, see
\cite{Platzer:2011dq, Hoeche:2011fd, Lonnblad:2012hz,
  Bellm:2016voq,Platzer:2018pmd} from some of these applications.

Especially when extending parton showers to include corrections beyond
leading order \cite{Li:2016yez,Hoche:2017iem,Dulat:2018vuy} as well as
sub-leading $\Nc$ effects, and in attempts to perform the evolution at
the amplitude level \cite{Platzer:2012np,Nagy:2012bt,Nagy:2015hwa,Platzer:2013fha,
  Isaacson:2018zdi,
  Platzer:2018pmd,Martinez:2018ffw,Forshaw:2019ver},
negative contributions to the splitting rates arise. These
contributions require the weighted Sudakov veto algorithm,
Alg.~\ref{algos:wei} in order to be included in the simulation.

\begin{algorithm}
  \begin{algorithmic}
    \STATE $Q' \gets Q, w \gets w_0$
    \LOOP
    \STATE A trial splitting scale and variables, $q,z$, are generated according to
    $S_R (q|Q',z,x)$, for example using Alg.~\ref{algos:sud}.
    \IF {$q=Q_0$}
    \STATE There is no emission and the cut-off scale $Q_0$ is returned
    while the event weight is kept at $w$.
    \ELSE
    \IF{{\bf rnd}$\le \epsilon$}
    \STATE The trial splitting variables $q,z$ are accepted, and
    \begin{equation}
      w\gets w \times \frac{1}{\epsilon} \times \frac{P(Q',z,x)}{R(Q',z,x)}\label{eq:accept}.
    \end{equation}
    \ELSE
    \STATE The emission is rejected, and the algorithm continues with
    \STATE
    \begin{eqnarray} 
      w&\gets& w\times \frac{1}{1-\epsilon}\times \left(1-\frac{P(q,z,x)}{R(q,z,x)}\right)\nonumber \\
      Q'&\gets& q
      \label{eq:reject}.
    \end{eqnarray}
    \ENDIF
    \ENDIF
    \ENDLOOP
  \end{algorithmic}
  \caption{\label{algos:wei} The weighted Sudakov veto algorithm from
    \cite{Bellm:2016voq}, starting with an initial non-zero weight
    $w_0$ and a scale $Q$. Here $S_R$ can be {\it any} density
    according to which we can generate branching variables, not
    necessarily defined by an overestimate of the target splitting
    kernel $P$. The acceptance probability, $0<\epsilon<1$ may in
    principle depend on the splitting variables, but for the purpose
    of this letter we will treat it as a constant, though this is not
    a conceptual limitation.}
\end{algorithm}

While the weighted Sudakov veto algorithm can be shown to correctly
account for negative contributions, an issue with this algorithm and
in general with weighting procedures, is largely varying weight
distributions which accumulate multiplicatively during the simulation of
an event, especially in presence of the competition algorithm.  In
\cite{Platzer:2018pmd} the weight degeneration was partly reduced by
only keeping event weights down to the winning scale in the
competition step. This greatly reduced the weight spreading, but did
not fully resolve the issue, in the sense that incorporating yet more
emissions with negative weights would have produced unmanageable
weights.

Approaching the parton shower precision era, we expect the issue of
negative or varying weights to be a severe obstacle that has to be
overcome, preferably by more general methods than to case by case monitor
weights and adjust weighting algorithms. We therefore suggest to
utilize the method of {\it resampling}
\cite{rubin:1987,gordon:salmond:smith:1993} in the context of Monte
Carlo event generators. We introduce the new method in this letter as
follows: In \secref{sec:Resampling}, we introduce the basics of the
resampling method, while a benchmark example, illustrating the
resampling power by comparing the proposed approach to a simplified
parton shower is given in \secref{sec:Benchmark}. We also show the
efficiency of resampling by taking a more realistic parton shower
\cite{CTEQ2015} with unit weights, destroying its statistical
convergence beyond recognition using the weighted Sudakov algorithm,
Alg.~\ref{algos:wei}, and recovering good statistical convergence
using the resampling method. Finally we make an outlook and discuss
improvements in \secref{sec:Conclusion}.

\section{Resampling}
\label{sec:Resampling}

The key idea behind {\it resampling} is to, rather than {\it
  reweigh}\footnote{This is typically referred to as 'unweighting' in
  the context of high energy physics event simulation.}  $N$ weighted
events, select $n$ events among the $N$ weighted ones in proportion to
their weights. This is done in an interleaved procedure which is
supposed to keep the weight distribution as narrow as possible at each
intermediate step. After each weight change (in our case induced by
the shower), all selected events are assigned the weight $N/n$ and are
further evolved separately.\footnote{The extension to an initially
  weighted sample of events (where the weight average is not
  necessarily one) is trivial: save the average event weight, and set
  (assuming $N$ kept events) each event weight to this average after
  finishing the simulation.}  It is important to mention that this
method will introduce correlated events; some of them will be
duplicated or have partly identical evolution history. The number $n$
of selected events may well be chosen equal to the number $N$ of
events, but can also be chosen differently.

Note that the resampling procedure will not alter the flexibility of
the simulation when it comes to predicting arbitrarily many
differential cross sections in one run of the program; in fact the
appearance of correlated events is also known in approaches using
Markov Chain Monte Carlo (MCMC) methods as explored for the hard
process cross sections in \cite{Kroeninger:2014bwa}.

In the more commonly used {\it reweighing} procedure, events with
small weights are selected with a low probability; however, if
selected, they are assigned the same nominal weight (typically $1$) as
other events.  Using
{\it resampling}, also the large weights are kept under
control. Instead of being allowed to increase without limit, all event
weights can be kept at unity through duplication of events with large
weights. This greatly improves the statistical convergence properties,
and mitigates the risks of obtaining distributions that appear statistically
wrong for a small number of events.

In an implementation of a parton shower, a natural way of implementing
the resampling procedure is to resample after each emission step, {\it
  i.e.} to let each event radiate according to whatever versions of
Sudakov factors and splitting kernels are used, and to then, among the
say $N'$ evolved events, pick $N'$ events in proportion to the weights
$w_i$, meaning that some events will be duplicated and some will be
forgotten. This can even be extended to apply to intermediate proposals
within the veto and competition algorithm itself; we will explored
both options in Sec.~\ref{sec:Benchmark}. Events which have reached
the cut-off scale $Q_0$ can be ignored in the resampling procedure,
since these are not evolved and will hence not acquire any further
weights in the shower evolution.  Rather, resampling among events
which already have unit weights can only impair the statistical
convergence, since some events will be ``forgotten'', whereas others
will be duplicated.

In the broader context of Monte Carlo simulation, resampling was first
introduced in \cite{rubin:1987,rubin:1988} as a means for transforming
Monte Carlo samples of weighted simulations into uniformly weighted
samples. Later, \cite{gordon:salmond:smith:1993} proposed resampling
as a tool for controlling weight degeneracy in \emph{sequential
  importance sampling methods} \cite{handschin:mayne:1969}. In
sequential importance sampling, weighted Monte Carlo samples targeting
sequences of probability distributions are formed by means of
recursive sampling and multiplicative weight updating operations. The
observation in \cite{gordon:salmond:smith:1993}, that recurrent
resampling guarantees the numerical stability of sequential importance
sampling estimators over large time horizons---a finding that was
later confirmed theoretically in \cite{delmoral:guionnet:2001}---lead
to an explosive interest in such \emph{sequential Monte Carlo methods}
during the last decades. Today sequential Monte Carlo methods
constitute a standard tool in the statistician's tool box and are
applied in a variety of scientific and engineering disciplines such as
computer vision, robotics, machine learning, automatic control,
image/signal processing, optimization, and finance; see e.g. the
collection \cite{doucet:defreitas:gordon:2001}.

We emphasize that when implemented properly, the 
resampling procedure does not involve any significant time-penalty. 
The selection of $n$ events among $N$ ones in proportion to their 
weights $w_i$ is equivalent to drawing $n$ independent and 
identically distributed indices $i_1, \ldots, i_n$ among 
$\{1, \ldots, N\}$ such that the probability that $i_\ell = j$ is $w_j / \sum_{\ell = 1}^N w_\ell$. 
Naively, one may, using the ``table-look-up'' method, generate 
each such index $i_\ell$ by simulating a uniform
(i.e., a uniformly distributed number) $u_\ell$ 
over $(0, 1)$, finding $k$ such that 
$$
\sum_{i = 1}^{k - 1} \frac{w_i}{\sum_{j = 1}^N w_j} < u_\ell \leq \sum_{i = 1}^k \frac{w_i}{\sum_{j = 1}^N w_j}, 
$$
and letting $i_\ell = k$. Assume that $n = N$ for simplicity; 
then, since using a simple binary tree search determining 
$k$ requires, on the average, $\log_2 N$ comparisons, 
the overall complexity for generating $\{ i_\ell \}$ is 
$N \log_2 N$. However, it is easily seen that if the 
uniforms  $\{ u_\ell \}$ are \emph{ordered}, 
$u_{(1)} \leq u_{(2)} \leq \ldots \leq u_{(N)}$, 
generating the $N$ indices $\{ i_\ell \}$ using 
the table-look-up method requires only $N$ 
comparisons. Hence, once a sample $\{ u_{(\ell)} \}$ of $N$ 
ordered uniforms can be simulated at a cost increasing 
only linearly in $N$, the overall complexity of the 
resampling procedure is linear in $N$. 

In the following we review one such simulation approach, which is
based on the observation that the distribution of the order statistics
$(u_{(1)}, \ldots, u_{(n)})$ of $n$ independent uniforms coincides
with the distribution of
$$
	\left(\prod_{i = 1}^n \sqrt[i]{v_i}, \prod_{i = 2}^n \sqrt[i]{v_i}, 
	\ldots, \sqrt[n - 1]{v_{n - 1}} \sqrt[n]{v_n}, \sqrt[n]{v_n} \right),
        $$
where the $n$ random variables $\{ v_i \}$ are 
again independent and uniformly distributed over $(0, 1)$; 
see \cite[Chapter~5]{devroye:1986}. 
As a consequence, a strategy 
for simulating  $\{ u_{(\ell)} \}$ is given by 
the following algorithm having indeed a linear 
complexity in $n$. 

\begin{algorithm}
  \begin{algorithmic}
    \STATE Generate $v_n$ from the uniform distribution over $(0, 1)$.
    \STATE Set $u_{(n)} = \sqrt[n]{v_n}$. 
    \FOR{$i = n - 1 \to 1$}
	    \STATE Generate $v_i$ from the uniform distribution over $(0, 1)$.
	    \STATE Set $u_{(i)} = u_{(i + 1)} \sqrt[i]{v_i}$.  
    \ENDFOR
  \end{algorithmic}
  \caption{\label{algos:prod:uniforms} Algorithm 
  generating a sample $\{ u_{(\ell)} \}$ of $n$ ordered uniforms. 
  The uniforms $\{ v_i \}$ are supposed to be independent.}
\end{algorithm}

Alternatively, one may apply the method of \emph{uniform spacings};
see again \cite[Chapter~5]{devroye:1986} for details.  A caveat of the
resampling procedure that one must be aware of is that it typically
leads to event path depletion for long simulations. Indeed, when the
number of iterations is large the
ancestral paths of the events will typically, if resampling is
performed systematically, coincide before a certain random point; in
other words, the event paths will form an ancestral tree with a common
ancestor.  For $n = N$ one may establish a bound on the expected
height of the ``crown'' of this ancestral tree, \emph{i.e.}, the
expected time distance from the last generation back to the most
recent common ancestor, which is proportional to $N \ln N$; see
\cite{jacob:murray:rubenthaler:2015}.  Thus, when the number of
iterations increases, the ratio of the height of the ``crown'' to that
of the ``trunk'' tends to zero, implying high variance of path space
Monte Carlo estimators.

The most trivial remedies for the path depletion phenomenon is to
increase the sample size $N$ or to reduce the number of resampling
operations. In \cite{kong:liu:wong:1994} it is suggested that
resampling should be triggered adaptively and only when the
\emph{coefficient of variation}

\begin{eqnarray}
	\CV^2 
	&=& N \sum_{i = 1}^N 
	\left( \frac{w_i}{\sum_{\ell = 1}^N w_\ell} - \frac{1}{N} \right)^2 
	\nonumber \\
	&=& N \sum_{i = 1}^N 
	\left( \frac{w_i}{\sum_{\ell = 1}^N w_\ell} \right)^2 - 1 
        \label{eq:CV}
\end{eqnarray}
exceeds a given threshold. The coefficient of 
variation detects weight skewness in the sense that 
$\CV^2$ is zero if all weights are equal. 
On the other hand, if the total weight is carried by a single 
draw (corresponding to the situation of maximal weight skewness), 
then $\CV^2$ is maximal and equal to $N - 1$. 
A closely related measure of weight skewness 
is the \emph{effective sample size}
$$
	\ESS = \frac{N}{1 +  \CV^2}. 
$$ 
Heuristically, $\ESS$ measures, as its 
name suggests, the number of samples that contribute 
effectively to the approximation. It takes on its maximal 
value $N$ when all the weights are equal and its minimal 
value $1$ when all weights are zero except for a single one. 
Thus, a possible adaptive resampling scheme could be 
obtained by executing resampling only when 
$\ESS$ falls below, say, $50\%$. We 
refer to \cite{cornebise:moulines:olsson:2008} for a 
discussion and a theoretical analysis of these measures 
in the context of sequential Monte Carlo sampling. 

With respect to negative contributions to the radiation
probability, showing up as negative weighted events, we note that
events may well keep a negative weight. In this case,
resampling is just carried through in proportion to $|w_i|$,
whereas events with $w_i<0$ contribute negatively when added to
histograms showing observables. Negative contributions,
for example from subleading color or NLO does thus not
pose an additional problem from the resampling perspective.

\section{Benchmark examples}
\label{sec:Benchmark}

In this section we illustrate the impact of the resampling procedure
both in a simplified setup and within a full parton shower
algorithm. As a toy model we consider a prototype shower algorithm
with infrared cutoff $Q_0$ and splitting kernels
\begin{equation}
  {\rm d}P(q,z,x) = a \frac{{\rm d}q}{q} \frac{(1+z^2){\rm d}z}{1-z}
  \theta(1-Q_0/q-z)\theta(z-x) \ ,
\end{equation}
with $a>0$ being some parameter (similar to a coupling constant), and $x$
an external parameter in the form of an initial ``momentum fraction''
(first arbitrarily set to 0.1), and later changing in the shower
with each splitting.

Starting from a scale $Q$ (initially we make the arbitrary choice
$Q=1$, and pick a cut-off value of $Q_0=0.01$) we obtain a new lower scale
$q$ and a momentum fraction $z$ (of
the previous momentum), sampled according to the Sudakov-type density
${\rm d}S_P(q|Q,z,x)$ associated with the splitting kernel above.

The momentum fraction parameter $x$ will be increased to $x/z$ after
the emission, such that for an emission characterized by
$(q_i,z_i,x_i)$ the variables for the next emission will be drawn
according to ${\rm d}S_P(q_{i+1}|q_i,z_{i+1},x_i/z_i=x_{i+1})$.  This
algorithm resembles ``backward evolution'' to some extent, but in this
case we rather want to define a model which features both the infrared
cutoff (upper bound, $1-\frac{Q_0}{Q}$, on $z$ and lower bound on
evolution scale $q>Q_0$) and the dynamically evolving phase space for
later emissions (lower bound on $z$), while still being a realistic
example of a parton shower, with double and single logarithmic
enhancement in $Q/q$.

\begin{figure*}
  \begin{center}
    \includegraphics[scale=0.6]{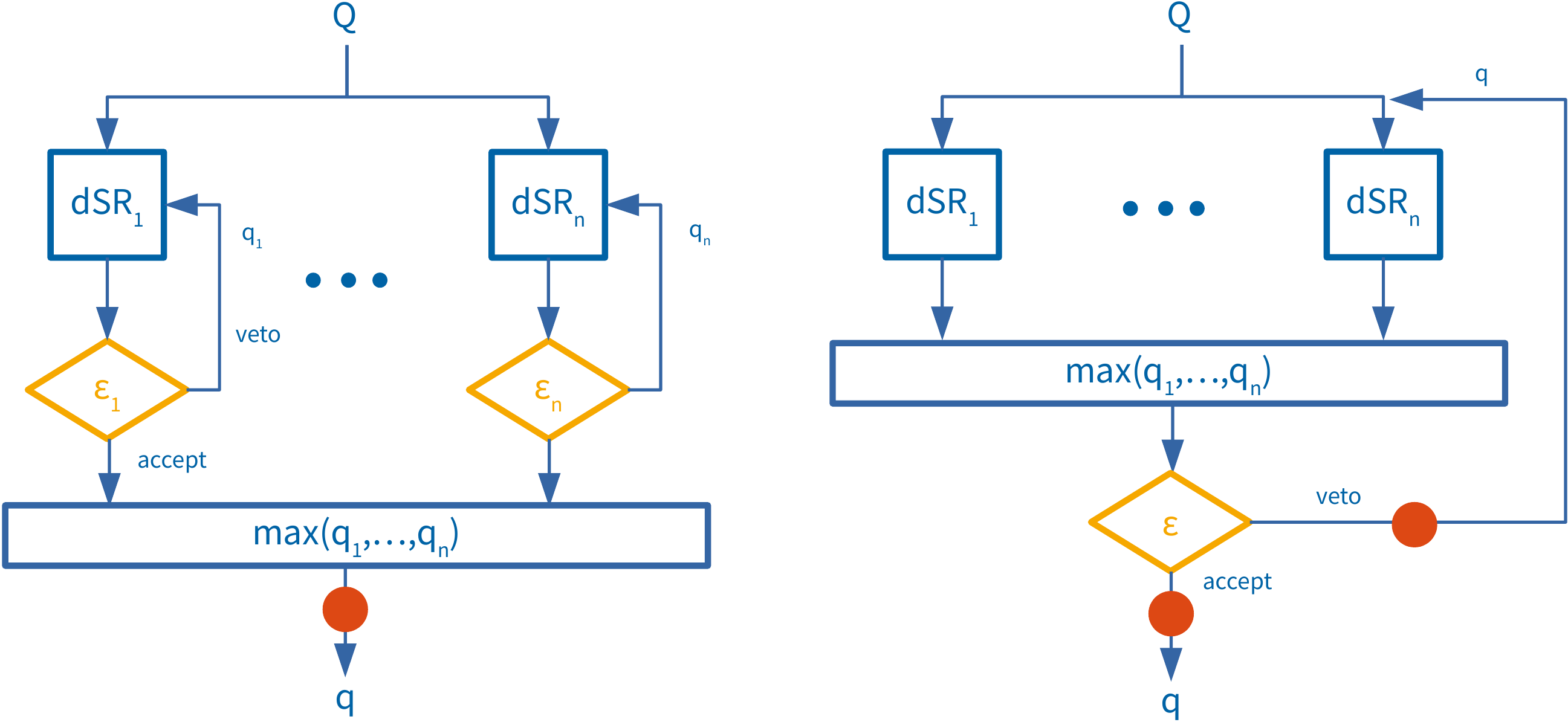}
  \end{center}  
  \caption{\label{fig:resamplingflow}Illustration of the different
    algorithms we consider as benchmark algorithms; we depict the
    algorithm as executed in each shower instance. Each ${\rm
      d}S_{R_i}$ block corresponds to a proposal density for one
    of $n$ competing channels, $\epsilon$ diamonds depict the
    acceptance/rejection step with the weighting procedure of the Sudakov
    veto algorithm, and the selection of the highest scale within the
    competition algorithm. Red dots indicate when the
    algorithm in each shower instance is interrupted and resampling in
    between the different showers is performed. Depending on the
    resampling strategy, events which reached the cutoff can be
    included in the resampling, or put aside as is done in our
    benchmark studies. The left flow diagram corresponds to the
    benchmark example discussed, while the right flow diagram
    corresponds to the implementation in the Python dipole
    shower. Re-entering after a veto step (the 'veto' branches), will
    start the proposal from the scale just vetoed.  Note that in the
    first case (left), resampling is performed only after emission (or
    shower termination), whereas in the second case, it is performed
    after each (sometimes rejected) emission attempt.  }
\end{figure*}

Proposals for the dynamic variables $q$ and $z$ are generated using a
splitting kernel of $R(q,z,x)=\frac{a}{q}\frac{2}{1-z}$ in place of
$P(q,z,x)=\frac{a}{q}\frac{1+z^2}{1-z}$, and with simplified phase space
boundaries $0<z <1-Q_0/q$.
We consider the distributions of the intermediate
variables $q$ and $z$, as well as the generated overall momentum fraction
$x$, for up to eight emissions.

To consider a realistic scenario, similar to what is encountered
in actual parton shower algorithms, where many emission
channels typically compete, we generate benchmark distributions by splitting
the parameter $a$ into a sum of different values,
(arbitrarily [0.01, 0.002, 0.003, 0.001, 0.01, 0.03, 0.002, 0.002 ,0.02, 0.02]) 
corresponding to competing channels.
The usage of weighted shower algorithms amplified by competition is
a major source of weight degeneration, as the weights from
the individually competing channels need to be
multiplied together.

\begin{figure}
  \begin{center}
    \includegraphics[width=0.45\textwidth]{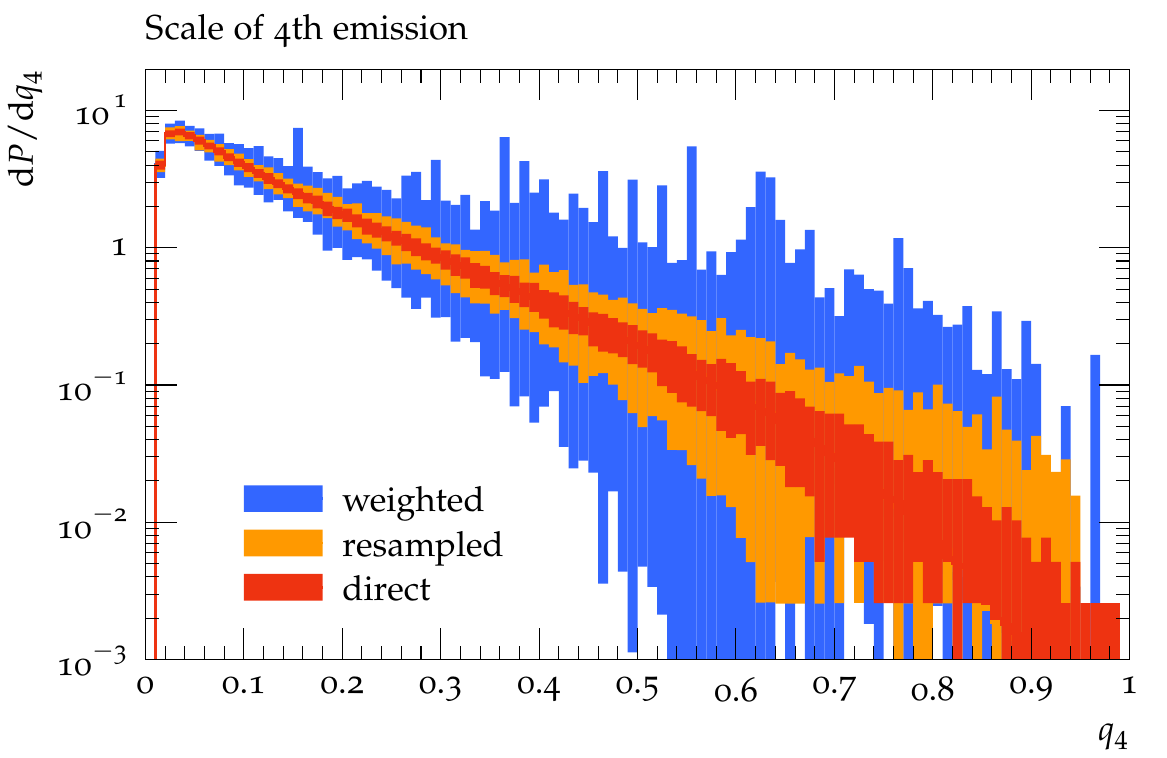}
  \end{center}  
  \caption{\label{fig:q4Distributions}Distribution of the scale of the
    $4^{\text{th}}$ emission, for
    sampling the distribution with the competition algorithm splitting
    $a$ into a sum across $10$ different competing channels, with red
    being the direct algorithm, blue the weighted Sudakov algorithm
    and orange the resampled, weighted Sudakov algorithm. The bands in
    the distribution reflect the minimum and maximum estimate
    encountered in $300$ runs with different random seeds and thus
    give a rough estimate of the width of the distribution of
    estimates.}
\end{figure}

We perform resampling after the weighted veto algorithm (choosing a
constant $\epsilon=0.5$ in this example) and the competition algorithm
have proposed a new transition, as schematically indicated in the left
panel in Fig.~\ref{fig:resamplingflow}.  In
Fig.~\ref{fig:q4Distributions} we illustrate the improvement of the
resampling algorithm.

We also try the resampling strategy out in a more realistic parton
shower setting, using the Python dipole parton shower from
\cite{CTEQ2015}.  In this case, we run 10\,000 events in parallel, and
each event is first assigned a new emission scale (or the termination
scale $Q_0$) using the standard overestimate Sudakov veto algorithm
combined with the competition algorithm. For each shower separately,
we then use one veto step as in Alg.~\ref{algos:wei} such that
some emissions are rejected. In the ratio $P/R$, $P$ contains the
actual splitting kernel and coupling constant, whereas $R$ contains an
overestimate with the singular part of the splitting kernel and an
overestimate of the strong coupling constant, see \cite{CTEQ2015} for
details. The constant $\epsilon$ in Alg.~\ref{algos:wei} is put
to 0.5.  This results in weighted shower events, and the resampling
step, where 10\,000 showers among the 10\,000 are randomly kept in
proportion to their weight, is performed, again resulting in
unweighted showers. This procedure is illustrated in the right panel
in \figref{fig:resamplingflow}, and we remark that the resampling step
can be combined with the shower in many different ways, for example,
as in the first example above (to the left in the figure) after each
actual emission (or reaching of $Q_0$), or, as in the second example
(to the right) also after rejected emission attempts.

\begin{figure}
  \centering
  \includegraphics[width=0.45\textwidth]{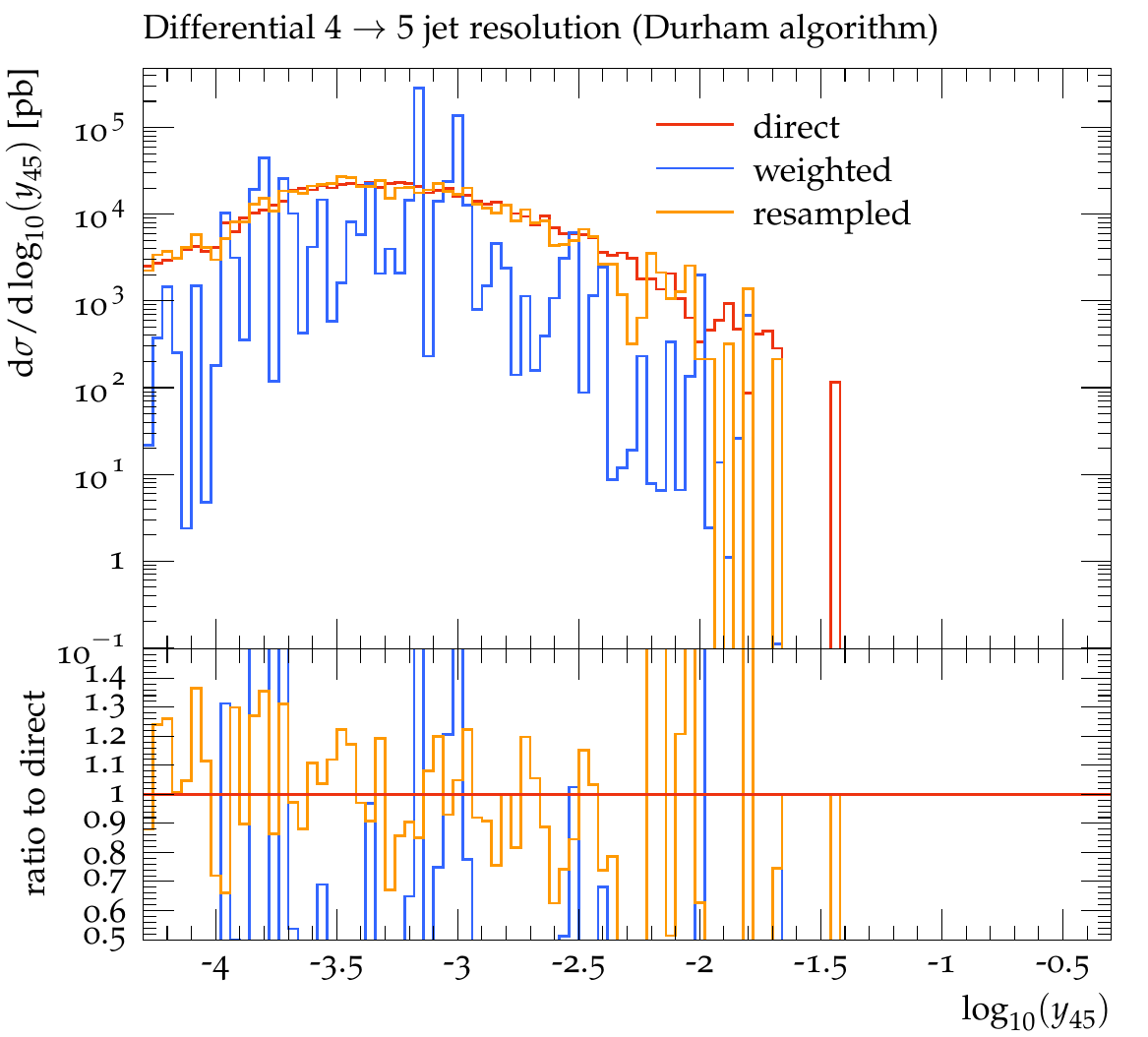}
  \caption{The Durham $y_{45}$ observable as implemented in
    \cite{CTEQ2015} (default). The same observable when sampled with
    Alg.~\ref{algos:wei} without resampling (no resampling),
    and when sampled with interleaved resampling after each emission
    or rejection step in Alg.~\ref{algos:wei}
    (resampling).} \label{fig:pythonshower}
\end{figure}

The result for the Durham $y_{45}$ jet observable \cite{Catani:1991hj}
is shown in \figref{fig:pythonshower}. The smooth curve in red gives
the unaltered result for the Durham $y_{45}$ jet observable
\cite{Catani:1991hj}; the jagged blue curve represents the same parton
shower with equally many events, but where the weighted Sudakov
algorithm, Alg.~\ref{algos:wei}, has been used.  The resulting
distribution is so jagged and distorted that it appears incompatible
at first sight. Only when adding interleaved resampling, in-between
the emissions steps (in orange) is it clear that the curves actually
represent the same statistical distribution. This beautifully
illustrates the power of the resampling method. To avoid oversampling,
the resampling is turned off for events for which the shower has
terminated.

We stress that, while the usage of the weighted Sudakov algorithm
in the above examples is completely unmotivated from a statistical
perspective, since the default starting curve already has optimal
statistics, the resampling procedure can equally well be applied in
scenario with real weight problems, coming for example from applying
the weighted Sudakov algorithm in a sub-leading $\Nc$ parton shower.

\section{Conclusion and outlook}
\label{sec:Conclusion}

In this note we advocate to use the resampling method for drastically
improving the statistical convergence for parton shower algorithms
where weighted algorithms are used.

The implementation tests, performed in the most simplistic way,
illustrate beautifully the power of the basic resampling method, and
prove its usefulness for parton showers. Nevertheless, we argue that
further improvement could be achievable by various extensions of the
algorithm to mitigate the ancestor depletion problem. One possible
approach is to furnish the proposed resampling-based algorithm with an
additional so-called \emph{backward simulation pass} generating
randomly new, non-collapsed ancestral paths on the stochastic grid
formed by the events generated by the algorithm
\cite{godsill:doucet:west:2004,douc:garivier:moulines:olsson:2009}. Another
direction of improvement goes in the direction of parallelization of
the algorithm, which is essential in scenarios with large sample
sizes. Parallelization of sequential Monte Carlo methods is
non-trivial due to the ``global'' sample interaction imposed by the
resampling operation. Nevertheless, a possibility is to divide the
full sample into a number of batches, or \emph{islands}, handled by
different processors, and to subject---``locally''---each island to
sequential weighing and resampling. Unfortunately, the division of the
sample into islands introduces additional bias, which may be of note
for a moderate number of islands. Following
\cite{verge:dubarry:delmoral:moulines:2015,delmoral:moulines:olsson:verge:2016},
this bias may be coped with by interleaving the evolution of the
distinct islands with \emph{island level selection} operations, in
which the islands are resampled according to their average weights.
    
At the more conceptual level, let us remark that the method we suggest
represents a first step in a change of paradigm, where parton showers
are viewed as tools for generating correct statistical distributions,
rather than tools for simulating independent events. We believe that
the most advanced and precise high energy physics simulations, if
they should run in an efficient way, will have no chance to avoid
weighted algorithms, and as such heavily need to rely on methods like
the resampling method outlined in this note. We will further
investigate the method, also including the possibility of making
cancellations in higher order corrections explicit already in
intermediate steps of such a Monte Carlo algorithm rather than by
adding large weights at the very end. We will also address the
structural challenges in implementing these methods in existing event
generators which we hope will help to make design decisions, keeping in
mind the necessity of methods like the one presented here.

\section*{Acknowledgments}

We thank Stefan Prestel for constructive feedback on the manuscript.

SP is grateful for the kind hospitality of Mainz Institute for
Theoretical Physics (MITP) of the DFG Cluster of Excellence PRISMA$^+$
(project ID 39083149), where some of this work has been carried out
and finalized.  JO and MS thank the Erwin Schr\"odinger Institute for
the kind hospitality during the PSR workshop. JO gratefully
acknowledges support by the Swedish Research Council, Grant
2018-05230.  MS was supported by the Swedish Research Council
(contract numbers 2012-02744 Torbjörn as well as the European Union's
Horizon 2020 research and innovation programme (grant agreement No
668679).

This work has also been supported in part by the European
Union’s Horizon 2020 research and innovation programme as part of the
Marie Skłodowska-Curie Innovative Training Network MCnetITN3 (grant
agreement no. 722104). SP acknowledges partial support by the COST
actions CA16201 ``PARTICLEFACE'' and CA16108 ``VBSCAN''.

\bibliography{resampling}

\end{document}